# TOWARDS THE SCALE INVARIANT COSMOLOGY

M.M. Verma*

Department of Physics,

Institute of Engineering and Technology,

CSJM University, Kanpur-208 024 (India)

**ABSTRACT :** An argument is made to show that the singularity in the General Theory of Relativity (GTR) is the expression of a non-Machian feature. It can be avoided with a scale-invariant dynamical theory, a property lacking in GTR. It is further argued that the global non-conservation of energy in GTR also results from the lack of scale-invariance and the field formulation presented by several authors can only resolve the problem in part. A truly scale-invariant theory is required to avoid these two problems in a more consistent approach.

* e-mail: sunilmmv@yahoo.com

## I. INTRODUCTION

A substantial amount of work has been focused over the past many years on solving the problems of singularity and global non-conservation of energy-momentum in General Theory of relativity (GTR). [Misner *et al* 1973 and further references therein]. The singularity is usually avoided by dropping one of the assumptions of the singularity theorems, using an approach motivated by the status of the total energy-density of matter in the universe [Senovilla 1990, Ruiz & Senovilla 1992, Stoeger *et al* 1995]. This approach often overlooks the basic element that the freedom from singularity must descend directly from the dynamical theory and not from the subjective dropping of such conditions or the choice of the metric. In section II it is shown that the more fundamental requirement is the scale-invariance which by itself makes a theory singularity-free.

Also, the global non-conservation of energy [Peebles 1993] in GTR is often explained by the field formulation [Grishchuk *et al* 1984]. In section III, it is discussed, however, as another fruitful dividend that the cause of this feature is embedded in the singularity of the theory which makes the time-axis contributions inhomogeneous, with the Lagrangian depending explicitly on the time.

A precise concluding statement is derived the end of section IV, acting as a fundamental argument in the development of consistent theory of cosmological models

## II. SCALE-INVARIANCE AND SINGULARITY

In the Quasi-Steady State Cosmology (QSSC) [Hoyle *et al*, 1994 a, b, 1995, Sachs *et al* 1996] with the equations

$$R_{ik} - \frac{1}{2}g_{ik}R = -\frac{6}{M^2}\left[T_{ik} - M_i M_k + \frac{1}{2}g_{ik}g^{pq}M_p M_q - \frac{g_{ik}}{6}\check{Z}M^2 + \frac{1}{6}M^2_{;ik}\right] \quad (1)$$

a scale change

$$\Omega(X) = M(X)/\overline{m} \quad (2)$$

may be introduced, where $\Omega(X)$ is a twice-differentiable function of the coordinates $X^i$. It is seen that in the Einstein conformal frame the scalar mass function remains constant, $\overline{m} = m_o$ (constant) in a spacetime manifold with the metric $\Omega^2 g_{ik}$, $0 < \Omega < \infty$, which reduces the above gravitational equations (1) to those of the GTR. The following observations are in order.

**(i)** It is mentioned that the singularity arises because of the occurrence of zero-mass hypersurfaces in the solutions of the QSSC equation [Hoyle & Narlikar 1974, Kembhavi 1979] leading to unphysical effects. However, if the new mass functions $\overline{m}$ = const. are also equal to zero all over, then $\Omega$ may still have a non-zero finite value at $M(X) = 0$ hypersurface, as required by the conformal transformation and a singularity may be averted. It may be noted that such a universe will be empty, like Milne's empty but singularity-free model with curvature parameter $k = -1$ and present deceleration parameter $q_0 = 0$. Eventually though, such theory becomes trivial and non-Machian.

**(ii)** If $M(X) = 0$ while $\overline{m}$ is still an undetermined, non-zero constant then reducing the scale to zero generates a singularity at time $t = 0$. Now such a singularity is the outcome of a non-Machian idea, i.e., $M(X) = 0$ with $\overline{m} \neq 0$ — a universe that is empty in one frame is not so in the other. From this , two probable explanations emerge. First, it appears that the singularity is the expression of the non-Machian character of the gravitational equations as those of the GTR. Secondly, when $M(X) = 0$, the measure of length scale (Compton wavelength $h/mc$) blows to infinity and this unphysical length scale results in singularity.

**(iii)** Since the QSSC equations are scale-invariant unlike those of the GTR, it seems pertinent from the above that a singularity-free theory is a necessary consequence of the scale-invariance, though the reverse is obviously not true, in view of some singularity-free solutions [Senovilla 1990, Ruiz & Senovilla 1992] for spatially inhomogeneous cosmological models. Though, these solutions show the complete causal curves with well-defined cylindrical symmetry, it is found, however, that these solutions do not satisfy the assumption of the compact trapped surfaces, among others in the Penrose-Hawking theorems [Hawking & Ellis 1973] for the exact perfect fluid ($\rho = 3p$).

In spherical models too, freedom from singularity can be achieved by the act of shear [Dadhich & Patel 1997], but it would be possible only if the instrumental role of shear in the collapse in the Raychaudhuri equations is surpassed by the counter-acceleration caused by it.

In the family of these models — either with cylindrical or spherical symmetry, the condition of the inhomogeneity of the spacetime (perturbed Friedman-Robertson-Walker metric) is indiscriminately used. However, this condition can be chosen *independently* of the dynamic theory (here GTR ) and the GTR is already not scale-invariant. Thus the avoidance of the singularity results from the 'choice' of the metric and not from the dynamical theory, while it must have actually descended from the later. Clearly therefore, if we drop the '*assumption*' of compact trapped surfaces which was motivated by the argument [Stoeger *et al* 1995] that energy density needed to thermalize the Cosmic Microwave Background Radiation (CMBR) is sufficient enough to converge the past geodesic congruence, (it is a circular reasoning!) we get no singularity in GTR. This means that the zero-mass hypersurfaces do not exist in other conformal frames.

# III. GLOBAL NON-CONSERVATION OF $T_{ik}$ IN GTR

It is known that the global conservation of energy is not obeyed in GTR (Peebles 1973, Misner *et al* 1973). If so, what in GTR is responsible for this global non-conservation ?

In the presence of gravitational fields, the expression

$$T^k_{i;k} = 0 \tag{3}$$

does not represent any conservation law (Landau & Lifshitz 1975) and so there it undergoes the transformation

$$T_{ik} \rightarrow (-g)(T_{ik} + t_{ik}) \tag{4}$$

where $t_{ik}$ is the energy momentum pseudo-tensor due to gravity whose energy is not localized in spacetime unlike electromagnetic field energy that can be fixed at a point of spacetime in all frames.

In GTR, however, $T_{ik}$ must *per se* incorporate the gravitational field contributions to energy and momentum. After all, $T_{ik}$ is generated by $L_{phys}$ in the action

$$A = \frac{1}{16\pi G} \int_V (R + 2\lambda)\sqrt{-g}\ d^4x + \int L_{phys}(X)\sqrt{-g}\ d^4x \tag{5}$$

whose variation with respect to a general Riemannian metric within a general spacetime volume $V$ gives the GTR equations. Hence,

$$T_{ik} = \frac{2}{\sqrt{-g}} \left[ \frac{\partial \sqrt{-g}}{\partial g^{ik}} L_{phys} - \frac{\partial}{\partial x^m} \frac{\partial(\sqrt{-g})}{\partial g^{ik}_{,m}} L_{phys} \right] \tag{6}$$

and gravity must be the 'in-built' character of matter and radiation (and any false vacuum, if it exists at all) that generate $T_{ik}$. But contrary to this, we find that we have to depend on the left hand side of the GTR equations

$$R_{ik} - \frac{1}{2} g_{ik} R = -8\pi G T_{ik} \tag{7}$$

for the $t_{ik}$ terms while any such contribution must be sitting with the 'sources' on the right hand side of these equations. It is unlike QSSC which gives

$$R_{ik} - \frac{1}{2} g_{ik} R = -8\pi G \left[ T_{ik} - \frac{2}{3}\left(C_i C_k - \frac{1}{4} g_{ik} C^l C_l\right) \right] \tag{8}$$

with the divergence of the right hand side being zero. In (8), under the condition of creative mode,

$$T^{ik}_{;k} \neq 0 \tag{9}$$

and for the non-creative mode,

$$T^{ik}_{;k} = 0 \tag{10).}$$

Here for the global conservation of the four-momentum, we need matter, electromagnetic radiation field and C-field (which may also be quantum vacuum-like) as the sources that incorporate gravity.

To solve this global non-conservation problem, although the field formulation of GTR has already been developed by several authors [Grishchuk *et al* 1984, Popova & Petrov 1988, Petrov 1993, Petrov & Narlikar 1996] having removed the pseudo-tensor $t_{ik}$ from the energy momentum tensor as its advantage, it now turns out that the global non-conservation of the four-momentum in GTR is the result of the initial singularity under the congruence of geodesics, put as an assumption in the Penrose-Hawking singularity theorems. This makes the overall time-axis contributions inhomogeneous and so the Largrangian depends explicitly on time so apparently at $t = 0$ creation event without any sources (in contrast to mini-bangs in the vast sea of C-field in QSSC).

## IV. CONCLUSION

It is interesting to note that, as discussed in section II, the singularity itself appears to be the consequence of the lack of scale-invariance of the dynamical theory for the evolution of the universe that consequently does not preserve the light cone structure globally. The well-known condition of congruence of geodesics resulting in gravitational collapse is obviously motivated by the argument of energy density with uncertain values, and when this condition is dropped, the singularity gets eliminated. This condition is thus trapped in an arbitrariness of argument.

The discussion of section III brings out the fact that if we have singularity-free dynamical equations of gravitation (such as in QSSC), the global conservation of four-momentum of matter plus field (including gravity) automatically follows.

Therefore, a simple argument can be represented as below through the following three conditions –

(a) Global preservation of light cone structure as required by a truly global theory *plus* scale - invariance of the theory.

(b) Singularity-free dynamical equations of gravity.

(c) Global conservation of four-momentum of matter *plus* field, including gravity.

The condition (a) leads to (b) and (b) leads to (c). Therefore the condition (c) ensues directly from (a) and bears out some fundamental strength of (a).

## V. ACKNOWLEDGMENT

The author thanks Dr. P. K. Sharma for his invaluable discussion on some of the arguments presented in this paper, and Dr. J. M. M. Senovilla for his encouraging remarks which motivated this work.